\documentclass[11pt,twoside]{article}
\input epsf
\usepackage{amstex, multicol, epsfig}
\begin{document}
\setlength{\parindent}{0em}
\setlength{\parskip}{1.5ex plus 0.5ex minus 0.5ex}
\title{Adaptive mesh refinement approach to construction of
initial data for black hole collisions.}
\renewcommand{\thefootnote}{\arabic{footnote}}
\footnotetext[1]{University of Texas at Austin, Austin, TX
78712}
\footnotetext[2]{Theoretical Astrophysics Center 
Juliane Maries Vej 30, 2100 Copenhagen, Denmark}
\footnotetext[3]{Code 6404, Naval Research Laboratory, Washington, DC 20375}
\footnotetext[4]{University Observatory, Juliane Maries Vej 30, 2100
Copenhagen, Denmark}
\footnotetext[5]{Astro Space Center of P.N. Lebedev
Physical Institute, Profsoyuznaya 84/32, Moscow, 117810, Russia} 
\footnotetext[6]{NORDITA, Blegdamsvej 17, 2100, Copenhagen, Denmark}
\author{\renewcommand{\thefootnote}{\arabic{footnote}}
Peter Diener \footnotemark[1] ,
Nina Jansen \footnotemark[2] ,  
Alexei Khokhlov \footnotemark[3] , 
Igor Novikov \footnotemark[2] {\footnotesize $^,$}\footnotemark[4]
{\footnotesize $^,$}\footnotemark[5] {\footnotesize $^,$}\footnotemark[6]}
\maketitle
\abstract{The initial data for black hole collisions is constructed using a 
conformal-imaging approach and a new adaptive mesh refinement technique, 
a fully threaded tree (FTT). We developed a second-order accurate
approach to the solution of the constraint equations on a non-uniformly 
refined high resolution Cartesian mesh including second-order accurate
treatment of boundary conditions at the black hole throats. Results of
test computations show convergence of the solution as the numerical 
resolution is increased. FTT-based mesh refinement reduces the
required memory and computer time by several orders of magnitude
compared to a uniform grid. This  opens up the possibility of using 
Cartesian meshes for very high resolution simulations of black hole 
collisions.}
\bigskip
\makeatletter
\renewcommand{\section}{\@startsection{section}{1}{0mm}
{-\baselineskip}{0.5\baselineskip}{\normalfont \normalsize \scshape}}
\renewcommand{\subsection}{\@startsection{subsection}{2}{0mm}
{-\baselineskip}{0.5\baselineskip}{\normalfont \normalsize \itshape}}
\makeatother
\begin{multicols}{2}
\section{\bf \normalfont \normalsize \scshape Introduction}
\label{sec:1}
\setlength{\parskip}{1.5ex plus 0.5ex minus 0.5ex}
This paper deals with the construction of initial data for black hole 
collisions on a high resolution Cartesian adaptive mesh. The problem
of black hole collisions is an important problem of the dynamics of
spacetime, and has applications to future observations of 
gravitational waves by gravitational observatories on Earth and in space.

The problem of black hole collisions is highly 
nonlinear and can only be solved numerically. 
A solution must be obtained within a large computational domain in
order to follow the outgoing gravitational waves far enough from the 
source. At the same time, very high resolution is required near the
black holes to describe the nonlinear dynamics of spacetime. 
Integration of the collision problem on a three-dimensional uniform
mesh requires enormous computational resources, and this remains one
of the major obstacles to obtaining an accurate solution.

Adaptive mesh refinement (AMR) can be used to overcome these problems
by introducing  high resolution only where and when it is required. AMR
is widely used in many  areas of computational physics and engineering.
It has been applied in a more limited way in general relativity 
\cite{bib:1}. There are several types of AMR. In a grid-based AMR, a 
hierarchy of grids is created, with finer grids overlayed on coarser
grids if a higher resolution is required \cite{bib:2}. An unstructured 
mesh  approach uses meshes consisting of cells of arbitrary shapes and
various sizes \cite{bib:3}. A cell-based approach to AMR uses 
rectangular meshes that are refined at the level of individual cells. 
This approach combines high accuracy of a regular mesh with
flexibility of unstructured AMR \cite{bib:4}. The new
introduced fully threaded tree (FTT) structure, which we use here,
leads to an efficient, massively parallel implementation of a
cell-based AMR \cite{bib:5}.

The first step in  solving the black hole collision problem is to 
construct the initial data. The widely used conformal-imaging approach
has been proposed in \cite{bib:6},\cite{bib:7},\cite{bib:8} and
developed in \cite{bib:9},\cite{bib:10}. Another approach to the 
construction of initial data was recently proposed in \cite{bib:11}.
Approaches for constructing initial data for certain specific cases of
black hole collisions were proposed in \cite{bib:12},\cite{bib:13},
\cite{bib:14}. The conformal-imaging approach
\cite{bib:6},\cite{bib:7},\cite{bib:8},\cite{bib:9},\cite{bib:10} 
consists of constructing the extrinsic curvature (momentum constraints
equations) using an imaging technique  and then solving a nonlinear 
elliptic equation for the conformal factor (energy constraint) with 
an appropriate mirror-image boundary condition. This approach is
adopted in this paper.

A  numerical technique for obtaining initial data for black hole 
collisions on a uniform Cartesian grid using conformal-imaging
approach is described in \cite{bib:10}. Two major problems with 
this approach mentioned in \cite{bib:10} are the low resolution of 
a uniform grid near black holes, and low-order accuracy and 
programming complexity of the inner boundary conditions at the black 
hole throats. In \cite{bib:10}, first-order accurate boundary 
conditions were implemented. The goal of this paper is to construct 
initial data for black hole collisions on a high resolution, Cartesian
FTT adaptive mesh. In the process of realizing this goal, we found 
that the accuracy of the solution critically depend on the accuracy of the
numerical implementation of the inner boundary condition. We developed
a simple second-order accurate algorithm for boundary conditions to
deal with this difficulty.
 
The paper is organized as follows. The next Section \ref{sec:2} 
presents the formulation of the problem and the equations solved. 
Section \ref{sec:3} describes the FTT technique, the finite-difference 
discretization of the problem, and the numerical solution 
techniques. Section \ref{sec:4} presents the results of the solutions for
various  configurations of two black holes configurations and compares
these with existing solutions.

\section{\bf \normalfont \normalsize \scshape 
Formulation of the problem}
\label{sec:2}
The ADM or 3+1 formulation  of the equations of general relativity
works with the metric  $\gamma^{ph}_{ij}$ and extrinsic curvature 
$K^{ph}_{ij}$ of three-dimensional spacelike hypersurfaces embedded in
the four-dimensional space-time, where $i,j = 1,2,3$, and the 
superscript $ph$ denotes the physical space. On the initial 
hypersurface, $\gamma^{ph}_{ij}$ and $K^{ph}_{ij}$ must satisfy the 
constraint equations \cite{bib:6}. The conformal-imaging approach 
assumes that the metric is conformally flat,
\begin{equation}
\label{eq:1}
   \gamma^{ph}_{ij} = \phi^4 \gamma_{ij}\quad,
\end{equation}
where $\gamma_{ij}$ is the metric of a background flat space.
This conformal transformation induces the corresponding transformation
of the extrinsic curvature
\begin{equation}
\label{eq:2}
          K^{ph}_{ij} = \phi^{-2} K_{ij} \quad.
\end{equation}
With the additional assumption of 
\begin{equation}
\label{eq:3}
         tr K = 0 \quad,                         
\end{equation}
the energy and momentum constraints are
\begin{equation}
\label{eq:4}
\nabla^2 \phi + {1\over 8} \phi^{-7} K_{ij} K^{ij} = 0 \quad, 
\end{equation}
and
\begin{equation}
\label{eq:5}
D_j K^{ij} = 0\quad,                            
\end{equation}
respectively, where $\nabla^2$ and $D_j$ are the laplacian and covariant 
derivative in flat space.

A solution to (\ref{eq:5}) for two  black holes with masses $M_\delta$,
linear momenta ${\bf P}_\delta$ and angular 
momenta ${\bf S}_\delta$, where  $\delta=1,2$ is the black hole index,
is \cite{bib:9}
\begin{equation}
\label{eq:6}
K_{ij}({\bf r}) = K^{lin}_{ij}({\bf r}) + K^{ang}_{ij}({\bf r})\quad,
\end{equation}
where
\begin{equation}
\label{eq:7}
\begin{split} 
K^{lin}_{ij}({\bf r}) &= 
3 \sum_{\delta=1}^2 \left(
{1\over 2R_\delta^2} \left( P_{\delta,i} n_{\delta,j} +
P_{\delta,j} n_{\delta,i} - \right. \right.\\ &\quad \left. \left. 
\left( \gamma_{ij} - n_{\delta,i} n_{\delta,j} \right)
P_{\delta,k} n_\delta^k \right)
\right)                                               
\end{split}
\end{equation}          
and
\begin{equation}
\label{eq:8}
\begin{split}    
K^{ang}_{ij}({\bf r}) &= 
3 \sum_{\delta=1}^2 \left(
{1\over R_\delta^3} 
\left( \epsilon_{kil} S_\delta^l n_\delta^k n_{\delta,j} +
\right. \right. \\
&\quad 
\left. \left. \epsilon_{kjl} S_\delta^l n_\delta^k n_{\delta,i} \right)
\right) \quad.
\end{split}
\end{equation}

In (\ref{eq:7}) and (\ref{eq:8}), the  comma in the subscripts
separates the index of a black hole from the coordinate component 
indices and is not a symbol of differentiation, $R_\delta=M_\delta/2$ 
is the black hole throat radius, and
${\bf n}_\delta = ( {\bf r} - {\bf r}_\delta ) 
/ \vert {\bf r} - {\bf r}_\delta \vert$ is the unit vector directed 
from the center of the $\delta$-th black hole ${\bf r}_\delta$
to the point $\bf r$. We work in units where $G=1,~c=1$. 
\begin{figure*}[ht]
\begin{center}
\epsfig{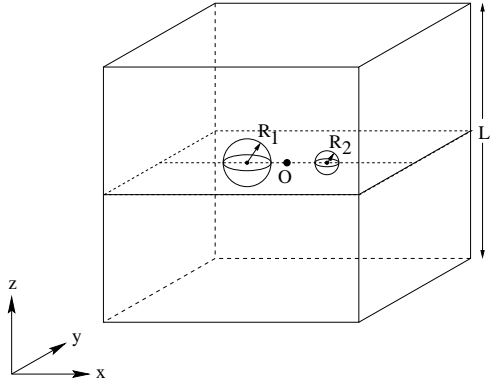}
\end{center}
\caption[]{\footnotesize This figure shows schematically the 
computational domain  used in the computations. The computational
domain is a cubic box of size $L$. Two black holes with 
throat radii $R_1$ and $R_2$ are positioned on the X axis in the
XY plane ($z=0$) at equal distances from the origin, O, of the 
coordinate system.}
\label{fig:1}
\end{figure*}

The inversion-symmetric solution to (\ref{eq:5}) can be obtained from 
(\ref{eq:6}) by applying an infinite series of mirror operators 
to (\ref{eq:6}), as described in \cite{bib:9}. Note, that before
applying the mirror operators to $ K^{ang}_{ij}$, this term must be 
divided by 2 since the image operators will double its value.
The series converges rapidly, and in practice only a few terms are 
taken. In this paper we take first five terms (for details see 
\cite{bib:15}). 

After the isometric solution for $K_{ij}$ is found, (\ref{eq:4}) must 
be solved subject to the isometry boundary condition at the black 
hole throats
\begin{equation}
\label{eq:9}
n_\delta^i D_i \phi = -{\phi \over 2 R_\delta}\quad,
\end{equation}
and the outer boundary condition $\phi \rightarrow 1$ when
$r \rightarrow \infty$. This boundary condition is represented by 
\cite{bib:9}
\begin{equation}
\label{eq:10}
{\partial \phi\over \partial r} = {1-\phi\over r}
\end{equation}
where $r$ is the distance from the center of the computational domain to the
boundary.
\section{\bf \normalfont \normalsize \scshape Numerical method}
\label{sec:3}
We work in the Cartesian coordinate system in the 
background flat space and find the solution 
within a cubic computational domain of size L. Figure \ref{fig:1}
shows the schematic representation of the computational domain. 
Centers of the throats of black holes are located in the XY plane 
$z=0$ on the line $y=0$ at equal distances from the origin. 
The space inside the black hole throats is cut out. The inner boundary
condition (\ref{eq:9}) is imposed at the surface of the two spheres, 
and the outer boundary condition is imposed at the border of the 
computational domain.
\begin{figure*}[ht]
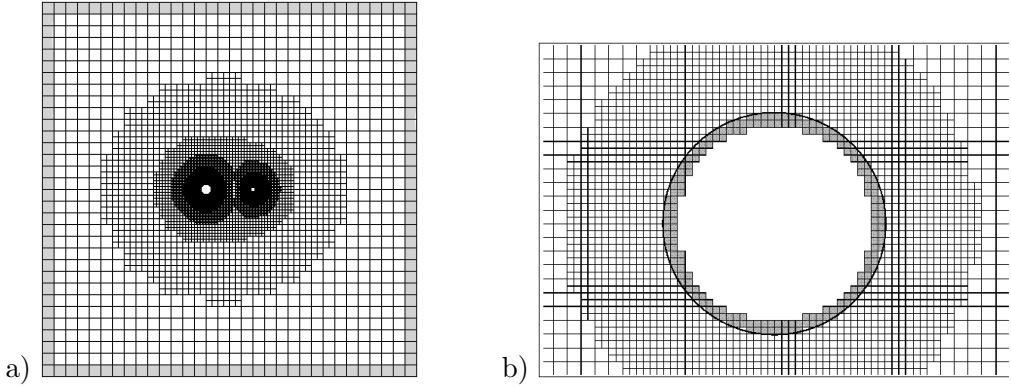

\begin{center}
a) \begin{minipage}[b]{6cm}
\epsfig{file=Figure2a.epsi, width=5cm}
\end{minipage}
b) \begin{minipage}[b]{7cm}
\epsfig{file=Figure2b.epsi, width=6cm}
\end{minipage}
\end{center}
\caption[]{\footnotesize A cut of a computational domain along the XY plane 
($z=0$) showing an adaptive mesh. The resolution increases near the
black holes. Internal cells are unshaded while boundary cells are
shown shaded.}
\label{fig:2}
\end{figure*}

Figure \ref{fig:2}
shows the cut of the computational domain through the $z=0$ plane and 
gives an example of an adaptive mesh used in computations. The values of all
variables are defined at cell centers. Figure \ref{fig:2} shows that
coarse cells are used at large distances from the black holes, and the
finest cells  are used near the throats where the gradient of $\phi$ 
is large. The grid is refined to achieve a desired accuracy of the 
solution as described below.

There are three types of cells. The first type are internal cells that are
actually used in computations (these cells are white in Figure 
\ref{fig:2}). The layer of boundary cells which is one cell wide
along the outer border of the computational domain is used to
define the outer boundary conditions. The layers of boundary 
cells inside the throats, two cells wide, are used to define the 
inner boundary conditions. A cell is considered as located inside 
a throat if its center is located inside. Boundary cells are indicated
as shaded on Figure \ref{fig:2}.

We solve (\ref{eq:4}) as follows. Similar to \cite{bib:11}, we introduce
a new unknown variable
\begin{equation}
\label{eq:11}
u = \phi - \alpha^{-1}\quad,
\end{equation}
where 
\begin{equation}
\label{eq:12}
\alpha^{-1} =  \sum_{\delta=1}^2 
\left( R_\delta\over\vert {\bf r} - {\bf r}_\delta \vert \right) \quad.
\end{equation}                                
The reason for using this transformation is that $u$ varies
slower than $\phi$ near the throats, and is more convenient for numerical 
calculations (see Section \ref{sec:41}). In Cartesian coordinates, 
(\ref{eq:4}) then becomes 
\begin{equation}
\label{eq:13}
\nabla^2 u + F(u) = 0\quad,
\end{equation}
where
\begin{equation}
\label{eq:14}
F(u) = { \beta \over \left( 1 + \alpha u \right)^7 }\quad,
\end{equation}
and 
\begin{equation}
\label{eq:15}
\beta = {1\over 8}\, \alpha^7 K_{ij} K^{ij}\quad.
\end{equation}

Equation (\ref{eq:13}) is a nonlinear elliptic equation. 
Below we describe the numerical 
procedure of finding its solution. First consider a cell of size $\Delta$
which has six neighbors of the same size. Let us number this cell and its 
neighbors with integers from 0 to 6, respectively. Then the discretized 
form of (\ref{eq:13}) is 
\begin{equation}
\label{eq:16}
{u_1 + u_2 + u_3 + u_4 + u_5 + u_6 -6 u_0 \over \Delta^2} + F(u_0) = 0
\end{equation}
A finite-difference form of (\ref{eq:13}) 
is more complicated for cells that have 
neighbors of different sizes and may involve larger number of neighbors 
in order to maintain second order accuracy. This is described in
Appendix \ref{ap:A}. 
In general, for every internal cell, the finite-difference discretization
may be written as
\begin{equation}
\label{eq:17}
f(u_0,u_1,u_2,..., u_n ) = 0 \quad ,
\end{equation}
where $u_1,...,u_n$ are the values of $u$ in $n$ neighboring points chosen 
to represent the finite-difference stencil of a cell. In the particular case
of a cell with all equal neighbors,
$f$ is defined by the left-hand side of (\ref{eq:16}).
\begin{figure*}[ht]
\begin{center}
\epsfig{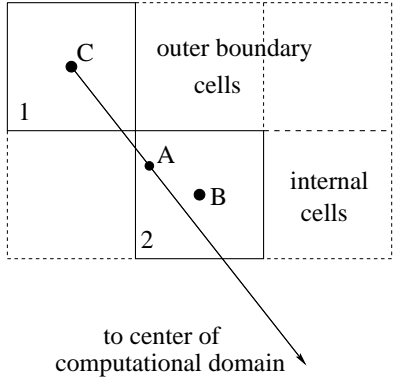}
\end{center}
\caption[]{\footnotesize An example of the computation of outer
boundary values.
Outer boundary cells are the three cells located at the top of the
figure. Cell 1 is an outer boundary cell, with center at point C.
Point A is the point located one cell size away from the C on the line
connecting point C with the center of the computational domain. Point
A is located inside cell 2, which is an internal cell centered on point B. 
The value of $\phi$ at point A, to be used in the outer boundary 
condition, is interpolated from point B with second order accuracy
using the derivatives evaluated at point B. These derivatives depend 
on $\phi$ at Cell 2 and it's neighboring cells, including outer 
boundary cells.}
\label{fig:3}
\end{figure*}

We solve the set of (\ref{eq:17}) by the Newton Gauss-Seidel method 
\cite{bib:16}, that is, we obtain a new guess of $u_0^{new}$ using
Newton iteration with respect to the unknown $ u_0$
\begin{equation}
\label{eq:18}
u_0^{new} = u_0 - f(u_0,...) 
\left( {\partial f(u_0,...)\over\partial u_0} \right)^{-1} \quad .
\end{equation}
Then we accelerate the convergence by using a successive
overrelaxation (SOR)
\begin{equation}
\label{eq:19}
u_0^{new} = \omega u_0^{new} + (1 - \omega ) u_0 \quad,
\end{equation}
where $\omega$  is the overrelaxation parameter.
For a simple case of all equal neighbors, (\ref{eq:18}) can be written as
\begin{equation}
\begin{split}
\label{eq:20}
u_0^{new} &= u_0 +\\
&{\left( u_1 + u_2 + u_3 + u_4 + u_5 + u_6 - 6 u_0 \right)\over 6 -
\Delta^2 \left(dF(u_0)\over du_0\right) } +\\
&\qquad {\left(\Delta^2 F(u_0)\right)
\over 6 - \Delta^2 \left(dF(u_0)\over du_0\right) } \quad .
\end{split}
\end{equation}
For stencils with non-equal neighbors the discretization of equation 
(\ref{eq:13}) is given in Appendix \ref{ap:A}, from which the
expressions for the Newton Gauss-Seidel iterations in those cases can 
be written explicitly. 

We select the value of $\omega$ from the interval $[1,\omega_{max}]$
by the method described in \cite{bib:17}. The value of $\omega_{max}$ is
initially set to  $\omega_{max} = 1.995$. 
If the solution begins to diverge during the iterations,
$\omega$ is reset to 1, $\omega_{max}$ is decreased by 2\%,
and the relaxation is continued allowing $\omega$ to increase up to the
new $\omega_{max}$ or until the solution starts to diverge again.
Iterations are terminated when 
\begin{equation}
\label{eq:21}
{u^{new}_0 - u_0 \over u_0} < \varepsilon
\end{equation}
for all $u_0$, where $\varepsilon$ is a predefined small number.
In this paper, we do not attempt to accelerate the iterations using, for
example, a multigrid or other sophisticated techniques since the initial
value problem must be solved only once.

The procedure described above assumes that the values of $u$ are known at
all neighbors. For internal cells that are close to a boundary, these
values are substituted with the values of $u$ in boundary cells. 
Now we describe the procedure of assigning values of $u$ to boundary cells. 
A similar technique was used for fluid flow simulations about complex 
bodies \cite{bib:18}.

Figure \ref{fig:3} illustrates the process for the outer boundary.
In this figure, cell 1 is a boundary cell. We need to define a new value 
$u^{new}_C$ in its center, point C. We find another point, A, which is 
located at a distance $\Delta$ (equal to the size of cell 1) from point $C$ 
along the line that connects $C$ with the center of the computational 
domain. The value of $u_A$ is found by the second order interpolation 
using the values of $u$ in cell 2 and all of its neighbors. The 
interpolation involves old values of $u$ in both internal cells and 
boundary cells. Then $\phi_A$ is computed using equation (\ref{eq:11})
with $\alpha^{-1}$ evaluated in point A. The finite-difference
expression of (\ref{eq:10}) can be written as
\begin{equation}
\label{eq:22}
{\phi^{new}_C - \phi_A\over\Delta} = 
{ 1 - ( \phi^{new}_C + \phi_A)/2 \over ( r_C - \Delta/2 ) } \quad ,
\end{equation}
and then solved for $\phi^{new}_C$. The value of $u^{new}_C$ is then 
finally found using equation (\ref{eq:11}) with $\alpha^{-1}$
calculated at point C.
\begin{figure*}[ht]
\begin{center}
\epsfig{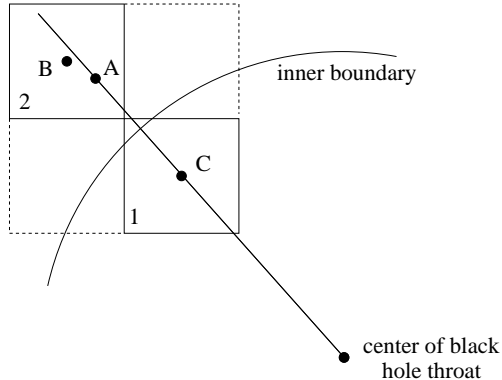}
\end{center}
\caption[]{\footnotesize An example of the computation of inner boundary
values. Cell 1, centered on point C, is the inner boundary cell.
Point A is the point located outside of the black hole throat on the line
between the center of the black hole throat and point C that has the
same distance to the boundary as point C. Point A is located inside cell 2, 
centered on point B. The value at point A, to be used in the inner
boundary condition, is interpolated from point B with second order
accuracy using the derivatives evaluated at point B. These derivatives
depend on the value of Cell 2 and it's neighboring cells, including
inner boundary cells.}
\label{fig:4}
\end{figure*}

Figure \ref{fig:4}
illustrates the process of defining $u$ for the inner boundary.
Cell 1 is located inside a throat, and we need to define a new value 
$u^{new}_C$ in its center, point C. The point $A$ is a point outside 
the throat that has the same distance to the inner boundary as $C$, 
along the normal to the throat passing through point C. Let us denote 
the distance between C and A as $\Delta_{AC}$. Again, the value of 
$u_A$ is found by second-order interpolation using old values of 
$u$ in cell 2 and its neighbors, and $\phi_A$ is calculated using 
equation (\ref{eq:11}). Then the boundary condition (\ref{eq:9}) becomes
\begin{equation}
\label{eq:23}
{ \phi^{new}_C - \phi_A  \over \Delta_{AC} } =
         { \phi_A + \phi^{new}_C \over 4R_{\delta} } \quad,
\end{equation}
which can be solved for $\phi^{new}_C$; $R_{\delta}$ is the radius of
the throat that contains point C. As before the value of $u^{new}_C$ is
then finally found using equation (\ref{eq:11}) with $\alpha^{-1}$ 
evaluated at point C.

After all inner and outer boundary points are defined, the next iteration
(\ref{eq:21})  is performed for internal cells and so on, until the 
iterations converge. The advantage of the method described above is 
that the new method applies the same numerical algorithm to all cells,
and that this algorithm is second-order accurate. The method described in 
\cite{bib:10} required  77 different numerical stencils corresponding to
different relative positions of the boundary and interior cells, 
and was only first-order accurate.

The computational domain used by the FTT is a cube of size $L$. It can be
subdivided to a number of cubic cells of various sizes 
$1/2,~1/4,~1/8,...$ of $L$. Cells are  organized in a tree with the 
direction of thread pointers inverted. These pointers are directed 
from children to neighbors of parent cells, as described in \cite{bib:5}.
The most important property of an FTT data structure is that 
all operations on it, including tree refinement and derefinement, 
can be performed in parallel. A computer memory overhead of FTT is 
extremely small: two integers per computational cell. All coding was 
done by using the FTTLIB software library \cite{bib:5} that contains 
functions for refinement, derefinement, finding neighbors, children, 
parents, coordinates of a cell, and performing parallel operations.

We characterize an FTT mesh by the minimum and maximum levels of 
leaves (unsplit cells) present in the tree, $l_{min}$ and $l_{max}$. 
We construct an adaptively refined mesh by starting with one computational
cell representing the entire computational domain and then by subsequently
subdividing it by a factor of two until we reach the level $l_{min}$.
We find a coarse solution for two black holes at level $l_{min}$ using
$u=1$ as an initial guess. After this solution is obtained, we
identify the regions that require more refined cells. These regions 
are then refined once, and a new converged solution is obtained. The 
old coarse solution is used as an initial guess for the new one. The 
procedure is repeated until the level of refinement reaches $l_{max}$.

\section{\bf \normalfont \normalsize \scshape Test results}
\label{sec:4}
\subsection{\bf \normalfont \normalsize 
\it One Schwarzschild black hole: one-dimensional test}
\label{sec:41}
Before presenting the results of three-dimensional test computations on an
FTT mesh, we will discuss the accuracy of our numerical method using a
simpler one-dimensional test problem. A single Schwarzschild black
hole at rest has an analytic solution for the conformal factor
\begin{equation}
\label{eq:24}
\phi = 1 + { R\over r}\quad,
\end{equation}
where $R$ is the throat radius and $r$ is the distance in the background
space from the center of the black hole throat. The one-dimensional
test was performed on a uniform grid in order to assess the influence
of our treatment of boundary conditions (\ref{eq:22}) and
(\ref{eq:23}), and the effects of changing the variable $\phi$ to a 
new variable $u$ (\ref{eq:11}) on the accuracy of the solution. In 
these calculations, we used a uniform one-dimensional grid consisting 
of $n=16$, $32$, and $64$ cells. Cells $2$ through $n-1$ were interior
cells. Cells $1$ and $n$ were the inner and outer boundary cells, 
respectively. The computations were performed for the grid size
$L=10$, throat radius $R=1$, and using convergence criterion
$\varepsilon= 6\times 10^{-15}$. The throat was located between the
first and second cell centers at a distance $\Delta r$ from the center
of the border cell~1. Three values of $\Delta r = 0.05\Delta$, 
$0.5\Delta$, and $0.95\Delta$ were considered, where $\Delta$ is the 
cell size. When $\Delta r = 0.5\Delta$, the throat is located exactly
in the middle between the points A and C in (\ref{eq:22}), and the
inner boundary condition (\ref{eq:22}) becomes second-order accurate 
regardless of the order of interpolation that is used for finding 
$u_A$. For $\Delta r = 0.05\Delta$ and $0.95\Delta$, the overall 
accuracy of the inner boundary condition depends on the interpolation 
used for finding $u_A$.

Numerical solutions were obtained using three different methods: 
{\it (a)} solving the finite-difference form of (\ref{eq:4}) for the 
original unknown variable $\phi$ and using {\it first order} 
interpolation to find $\phi_A$ (the rest of the boundary condition 
procedure was identical to that described in Section \ref{sec:2}),
{\it (b)}  solving for the original variable $\phi$ but using second order 
interpolation, and finally, {\it (c)} using both the second order 
interpolation and solving for the new unknown $u$ as described in 
Section \ref{sec:2}.
\begin{table*}[ht]
\begin{center}
{\itshape Accuracy of one-dimensional computations.}\vspace{0.2cm}\\
\begin{tabular}{|lllll|} \hline 
Method& $\Delta r / \Delta$ & n= 16 & n=32 & n=64 \\ \hline
& 0.95    & $-2.9\times 10^{-1}$ &  $-7.2\times 10^{-2}$  & 
$-1.8\times 10^{-2}$ \\
(a)&0.5   & $-6.1\times 10^{-2}$ &  $-1.7\times 10^{-2}$  &
$-5.0\times 10^{-3}$\\
&0.05  & $-3.8\times 10^{-1}$ &  $-4.6\times 10^{-1}$  & 
$-4.7\times 10^{-1}$ \\ \hline
&0.95  & $-2.9\times 10^{-1}$ &  $-7.0\times 10^{-2}$  & $-1.7\times
10^{-2}$ \\
(b)&0.5   & $-6.1\times 10^{-2}$ &  $-1.7\times 10^{-2}$  &
$-5.0\times 10^{-3}$ \\
&0.05  & $-3.0\times 10^{-1}$ &  $-1.2\times 10^{-1}$  & $-4.2\times
10^{-2}$ \\ \hline
&0.95  & $-3.4(3.5)\times 10^{-1}$ &  $-8.7(8.8)\times 10^{-2}$  &
$-2.2(2.2)\times 10^{-2}$ \\
(c)&0.5   & $-7.4(9.8)\times 10^{-2}$ &  $-2.1(2.4)\times 10^{-2}$  &
$-5.7(6.1)\times 10^{-3}$ \\
&0.05  & $-9.9(9.8)\times 10^{-4}$ &  $-1.4(2.4)\times 10^{-4}$  &
$-5.7(6.1)\times 10^{-5}$ \\ \hline
\end{tabular}
\end{center}
\caption[]{\footnotesize Relative accuracy of the numerical solutions of
(\ref{eq:4}) for a Schwarzschild black hole obtained using three 
different methods ({\it a,b,c}), different resolutions ($n$), and 
different location of the throat relative to a grid ($\Delta r$). 
Numbers in parentheses for the method ({\it c}) is the accuracy 
estimate based on the consideration of the error introduced by
the inner boundary condition (see Section \ref{sec:41}).}
\label{tab:1}
\end{table*}
Table \ref{tab:1} compares the numerical and analytical solutions for these
three cases by showing the maximum relative deviation of the numerical
solution from the analytical solution for the interior points of the 
grid. As can be seen from Table \ref{tab:1}, the accuracy varies with
the grid resolution ($n$), the method of interpolation, and the choice
of the unknown variable ($\phi$ or $u$). It also depends on the exact 
location of the throat relative to grid points ($\Delta r/\Delta$). 
As we expect various relative locations of the throats relative to 
grid points in three-dimensional calculations, we need a numerical 
procedure that provides a second-order accuracy in all cases.

Results using method {\it (a)} show that the accuracy of 
the solution using first order interpolation for $\phi_A$ 
is unacceptable. The third row of Table \ref{tab:1} shows 
that the accuracy does not improve with increasing number of cells. 
Results using method {\it (b)}  show that second order interpolation 
for $\phi_A$ (method {\it (b)}) leads to the overall second-order
algorithm. The accuracy of the solution increases roughly by a factor 
of four when the grid resolution is doubled. Results for method 
{\it (c)} shows that the accuracy is further improved for small $\Delta r$. 

It is possible to give an analytical estimate for an error 
introduced by the numerical inner boundary condition (\ref{eq:22}) for the
Schwarzschild black hole case. In this case, the general solution of 
(\ref{eq:4}), limited at infinity, is $\phi = c + b/r$. Let us assume 
that the numerical outer boundary condition does not introduce any
error, and the solution in interior points is found exactly. Thus, 
the numerical approximation of the inner boundary condition is a 
unique source of numerical error. Then the numerical solution would have 
the form $\phi = 1 + b/r$. The difference between $b$ and $R$ in 
(\ref{eq:24}) then will determine the overall error in the solution. 
We can find $b$ by substituting $\phi = 1 + b/r$ into (\ref{eq:22}). 
The estimate of the relative error then is
\begin{equation}
\label{eq:25}
{\rm Relative~error} = {R-b\over R} = \left(\Delta r \over 2
R\right)^2 \quad .
\end{equation}
The estimate of the error using (\ref{eq:25}) is given in Table 
\ref{tab:1} in brackets for the method {\it (c)}. The comparison with 
the numerical error indicates that for method {\it (c)}, the error in 
the solution is second-order and is determined by the accuracy of the 
inner boundary condition rather than by errors of numerical
calculations for internal cells. 

\subsection{Time-symmetric initial data for two black holes.}
\label{sec:42}
Next, we consider the case of two black holes with
${\bf P}_\delta = {\bf S}_\delta =0$ that have masses $M_1 = 1$,
$M_2=2$ and located (positions of the centers of their throats) at 
${\bf r}_1 = (-4,0,0)$, ${\bf r}_2 = (4,0,0)$ with finite separation 
$\vert {\bf r}_1 - {\bf r}_2 \vert = 8$. The size of the computational
domain is $L=64$. Numerical solutions were obtained using FTT
adaptive meshes with different increasing resolutions near the black
hole throats. We characterize the resolution by specifying the minimum
and maximum levels of cells in the tree, $l_{min}$ and $l_{max}$. 
The cell size at a given level $l$ is $\Delta_l = L \cdot 2^{-l}$.
The computations were performed on meshes with $l_{min} = 4,5,6$ and 
$l_{max} = 6,7,8,9,10,11$. The refinement criterion for this case was 
the requirement that
\begin{equation}
\begin{split}
\label{eq:26}
\eta &= {\Delta \over \phi^4} \left(
     \left(\partial\phi^4\over\partial x\right)^2 + 
     \left(\partial\phi^4\over\partial y\right)^2 +
     \left(\partial\phi^4\over\partial z\right)^2
                             \right)^{1/2} \\
&< 0.05
\end{split}
\end{equation}
in every cell, where partial derivatives in (\ref{eq:26}) are determined by 
the numerical differentiation. We used $\varepsilon = 6\times 10^{-7}$
in (\ref{eq:21}) to terminate iterations.

The mirror-image symmetric analytic solution for two time-symmetric 
black holes is given in Appendix \ref{ap:B} (for details of derivation
see \cite{bib:15}). Table \ref{tab:2} gives the comparison
of the numerical solutions with fixed $l_{min} = 5$ and varying 
$l_{max} = 5-11$ with the analytical solution. The table shows the 
maximum deviation of a numerical solution from the analytical one. 
It also shows the level of a cell where the maximum error was 
found. From the table we see that the accuracy of the solution 
increases approximately linearly with increasing $l_{max}$, and that the 
maximum error is located at maximum level of refinement  near the throats.
When we compare solutions obtained on different meshes on which the 
resolution was increased on all levels simultaneously 
(Table \ref{tab:3}), we observe better than linear convergence, as 
it should be expected.

The computations performed on an adaptive mesh allow us to save a
significant amount of computational resources. For example, our solution 
obtained on the $l_{min} = 5$, $l_{max} = 11$ adaptive mesh
used $6\times 10^5$ computational cells. An equivalent uniform-grid
computation with the same resolution  near the throats would have required
using a $2048^3$ uniform Cartesian grid with $\simeq 8\times 10^9$
cells. That is, in this case the computational gain was $\sim 10^4$.
\begin{table*}
\begin{center}
{\itshape Accuracy of two black 
hole time-symmetric computations.}\vspace{0.2cm}\\
\begin{tabular}{|cll|} \hline 
$l_{min}~-~l_{max}$ & Max. error & $l_{err}$ \\ \hline
5 - 5 & $3.3\times 10^{-2}$ & 5 \\
5 - 6 & $4.8\times 10^{-2}$ & 6 \\
5 - 7 & $2.9\times 10^{-2}$ & 7 \\
5 - 8 & $1.0\times 10^{-2}$ & 8 \\
5 - 9 & $1.2\times 10^{-2}$ & 9 \\
5 - 10 & $3.9\times 10^{-3}$ & 10 \\
5 - 11 & $8.9\times 10^{-4}$ & 11 \\ \hline
\end{tabular}
\end{center}
\caption[]{\footnotesize The table shows the maximum relative
deviation of the 
numerical solutions for the two Schwarzschild black holes 
(Section \ref{sec:42}), and the level of cells $l_{err}$ where the 
maximum error is located for computations with different maximum 
resolution.}
\label{tab:2}
\end{table*}
\begin{table*}
\begin{center}
{\itshape Accuracy of two black hole time-symmetric 
computations.}\vspace{0.2cm} \\
\begin{tabular}{|cl|} \hline 
$l_{min}~-~l_{max}$ & Max. error \\ \hline
4 - 8 & $3.6\times 10^{-2}$  \\
5 - 9 & $1.2\times 10^{-2}$ \\
6 - 10 & $3.7\times 10^{-3}$ \\ \hline
\end{tabular}
\end{center}
\caption[]{\footnotesize The table shows the maximum relative
deviation of the
numerical solutions for the two Schwarzschild black holes 
(Section \ref{sec:42}) for computations where both minimum and maximum
resolutions were increased simultaneously.}
\label{tab:3}
\end{table*}
\begin{table*}
\begin{center}
{\itshape Accuracy of computations of two black hole with non-zero linear and
angular momenta.}\vspace{0.2cm}\\
\begin{tabular}{|cll|} \hline 
$l_{min}~-~l_{max}$ & Max. error & Avg. error \\ \hline
5 - 5 & $1.1\times 10^{-1}$ & $1.1\times 10^{-2}$ \\
5 - 6 & $6.5\times 10^{-2}$ & $9.0\times 10^{-3}$ \\
5 - 7 & $1.6\times 10^{-2}$ & $1.8\times 10^{-3}$ \\
5 - 8 & $1.7\times 10^{-2}$ & $2.1\times 10^{-3}$ \\
5 - 9 & $5.3\times 10^{-3}$ & $6.4\times 10^{-4}$ \\
5 - 10 & $7.5\times 10^{-4}$ & $6.7\times 10^{-5}$ \\ \hline
\end{tabular}
\end{center}
\caption[]{\footnotesize Comparison of coarse solutions to the finest 
solution on the $l_{min}=5$, $l_{max}=11$ mesh. The table gives the maximum
relative deviation (Max. error) and the average relative deviation 
(Avg. error) of the numerical solutions computed in Section \ref{sec:43}.}
\label{tab:4}
\end{table*}
\subsection{Two black holes with linear and angular momenta}
\label{sec:43}
Cook et al. \cite{bib:10} considered the initial conditions for 
two black holes of equal mass $M_1=M_2=2$ with non-zero linear and angular
momenta (case A1B8 in Table 3 of \cite{bib:10}). In our coordinate system 
(Figure \ref{fig:1}), the components of linear and angular momenta of 
the black holes
for the case A18B are ${\bf P}_1=-{\bf P}_2=(0,0,-14)$, and ${\bf
S}_1=(280,280,0)$ and ${\bf S}_2=(0,280,280)$, respectively. 
The throats are located at ${\bf r}_1 = (4,0,0)$ and 
${\bf r}_2 = (-4,0,0)$ with a relative separation equal eight. 
We computed the A1B8 case using the same size of the computational 
domain, $L=28.8$ as that used in \cite{bib:10}, and using
a series of refined meshes with increasing resolution near the throats,
$l_{min} = 5$,  $l_{max} = 5-11$. We used the same  value of $\varepsilon$
as in Section \ref{sec:42} but used a modified mesh refinement criterion
\begin{equation}
\label{eq:27}
\begin{split}
\eta &= \max \left(
     {\Delta \over \phi^4} \left(
     \left(\partial\phi^4\over\partial x\right)^2 + 
     \left(\partial\phi^4\over\partial y\right)^2 + \right.\right. \\
&\qquad \left.\left.
     \left(\partial\phi^4\over\partial z\right)^2
                             \right)^{1/2}, \vert K_{ij}\vert \right)~< 0.05
\end{split}
\end{equation}

Table \ref{tab:4} compares coarse solutions $l_{min}=5$,
$l_{max}=5-10$ with the finest solution obtained on the  $l_{min}=5$, 
$l_{max}=11$ mesh. It shows that both the maximum and the average 
deviation of the solutions decreases by more than two orders of 
magnitude when the maximum resolution near the throats is increased 
by a factor of 32. The solutions in \cite{bib:10} did not show 
improvement with increasing resolution (see their Table \ref{tab:3}). 
\begin{figure*}[ht]
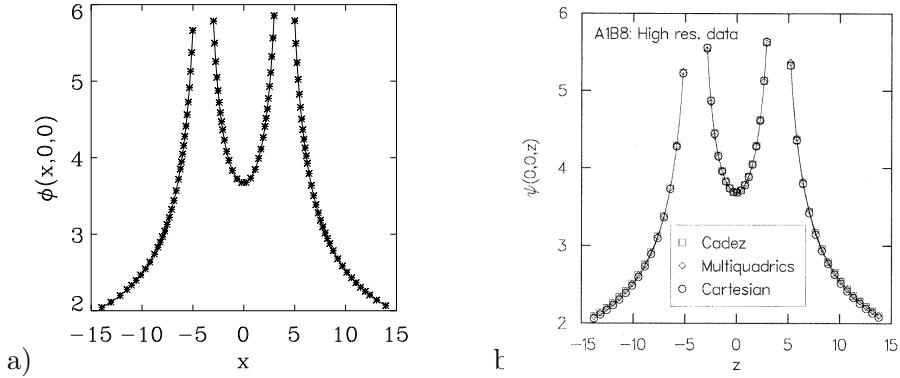

\begin{center}
a)\begin{minipage}[b]{6cm}
\epsfig{file=figure5a.epsi, width=5cm}
\end{minipage}
b)\begin{minipage}[b]{6cm}
\epsfig{file=figure5b.epsi, width=5cm}
\end{minipage}
\end{center}
\caption[]{\footnotesize A comparison of $\phi$ for the A1B8 case 
presented in
\cite{bib:10} (their Figure 4.) (b) and our computations with resolution 
$l_{min}=5$ and $l_{max}=8$ (a). Differences are not visible on the
scale of these graphs, except for the fact that we have better 
resolution near the holes which is why we have higher values close
to the holes.}
\label{fig:5}
\end{figure*}
Figure 5 shows the comparison of $\phi$ on the line passing through
the centers of the throats computed in this paper with resolution 
$l_{min}=5$, $l_{max}=8$ with the results presented in \cite{bib:10} 
(their Figure 4). In our computations, finer cells 
cluster near the throats where the gradient in the solution is larger,
whereas in \cite{bib:10}, cells have the same size and are uniformly 
distributed in space. This combined with the overall second-order 
accuracy of our method is the reason why the adaptive mesh refinement 
solution improves when the mesh is refined (see Table \ref{tab:4}). 
\section{Conclusions.}
\label{sec:5}
In this paper, we applied a new adaptive mesh refinement technique, a 
fully threaded tree (FTT), for the construction of initial data for
the problem of the collision of two black holes. FTT allows mesh to be
refined on the level of individual cells and leads to efficient 
computational algorithms. Adaptive mesh refinement is very important
to the problem of black hole collisions because a very high 
resolution is required  for obtaining an accurate solution.

We have developed a second-order approach to representing both 
the inner boundary conditions at the throats of black holes and the
outer boundary conditions. This allowed us to implement an approach to the
solution of the energy constraint that is formally second-order accurate.
We presented results of tests for two black holes that demonstrated a 
good improvement of the accuracy of the solution when the numerical 
resolution was increased. 

The FTT-based AMR approach gives a gain of several orders of magnitude
in savings of both memory and computer time (Section \ref{sec:42}),
and opens up the possibility of using Cartesian meshes for very high
resolution simulations of black hole collisions. A second-order 
boundary condition technique similar to that developed in this paper 
can be applied for the integration of initial conditions in time. We 
plan to use these techniques for time integration of the black hole 
collision problem.

{\it Acknowledgments.} 
This paper is based in part on the Master Thesis of Ms. Nina Jansen 
\cite{bib:15}. We thank J. Craig Wheeler, Elaine S. Oran and Jay P. 
Boris for their support, encouragement, and discussions, and 
Almadena Yu. Chtchelkanova for help with FTT. The work was supported 
in part by the NSF grant AST-94-17083, Office
of Naval Research, DARPA, Danish Natural Science Research Council
grant No.9401635, and by Danmarks Grundforskningsfond 
through its support for the establishment of the Theoretical 
Astrophysics Center.
\appendix
\section{Appendix: Finite-difference formulas on the Fully
Threaded Tree}
\label{ap:A}
\begin{figure*}[ht]
\begin{center}
a)\begin{minipage}[b]{5cm}
\epsfig{file=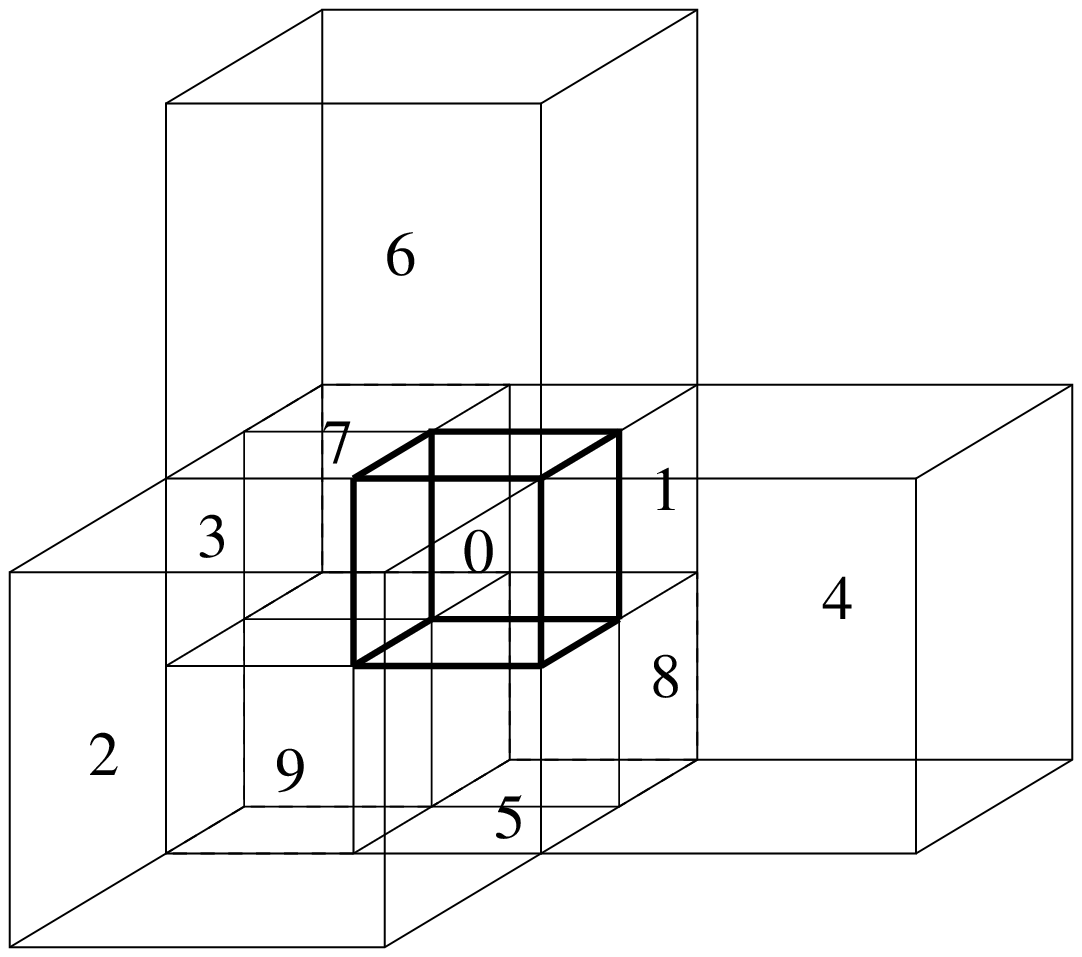,width=4.0cm, angle=270}
\end{minipage}
b)\begin{minipage}[b]{5.0cm}
\epsfig{file=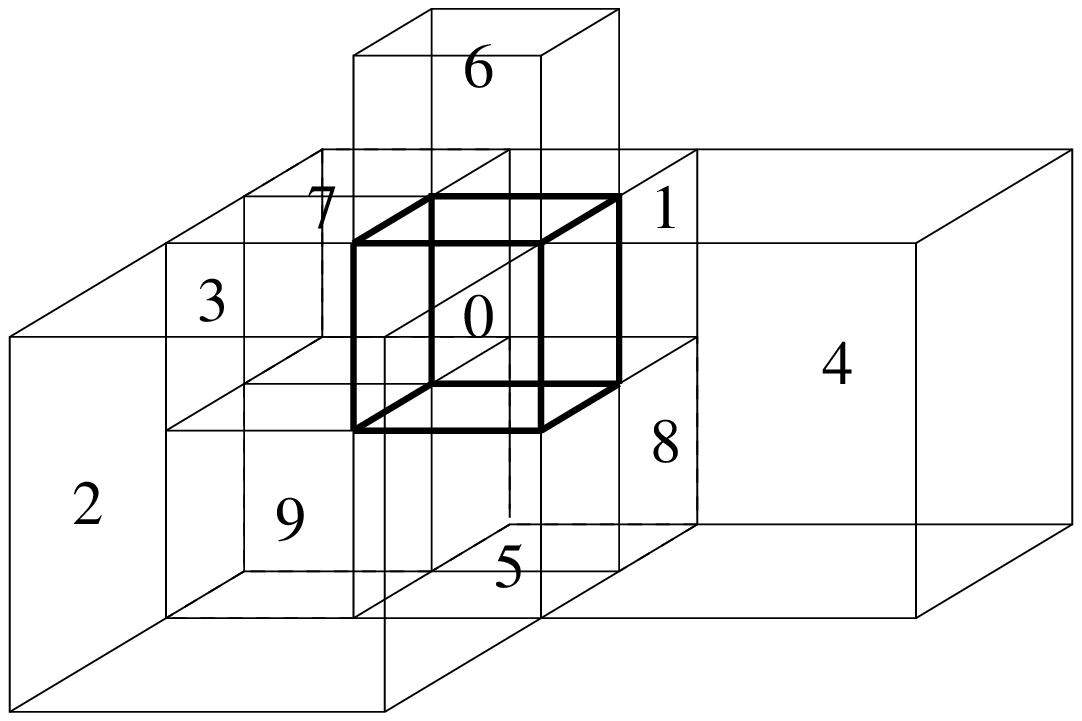,width=4.0cm,angle=270}
\end{minipage}
c)\begin{minipage}[b]{6.0cm}
\epsfig{file=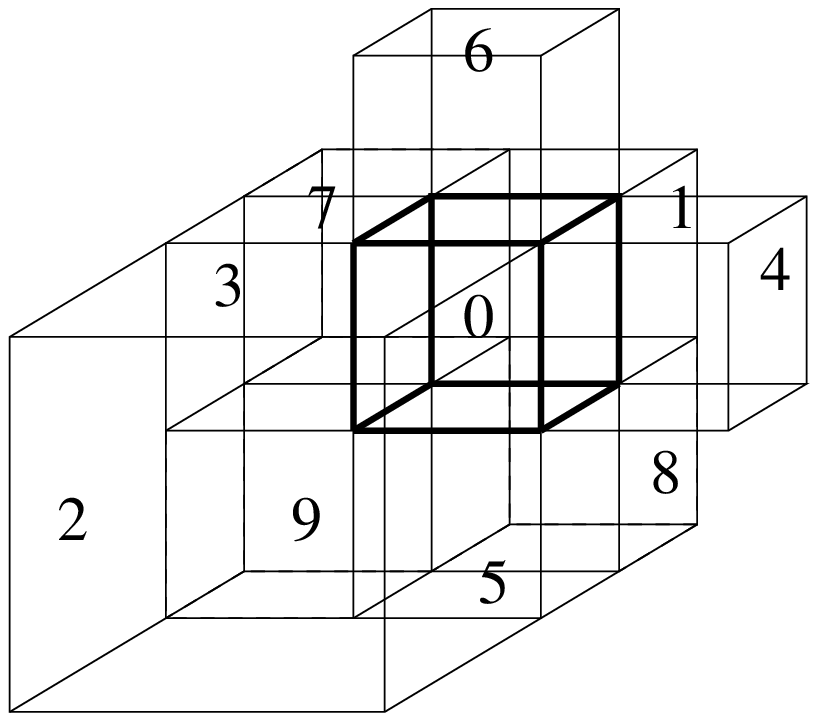,width=4.0cm,angle=270}
\end{minipage}
d)\begin{minipage}[b]{6.0cm}
\epsfig{file=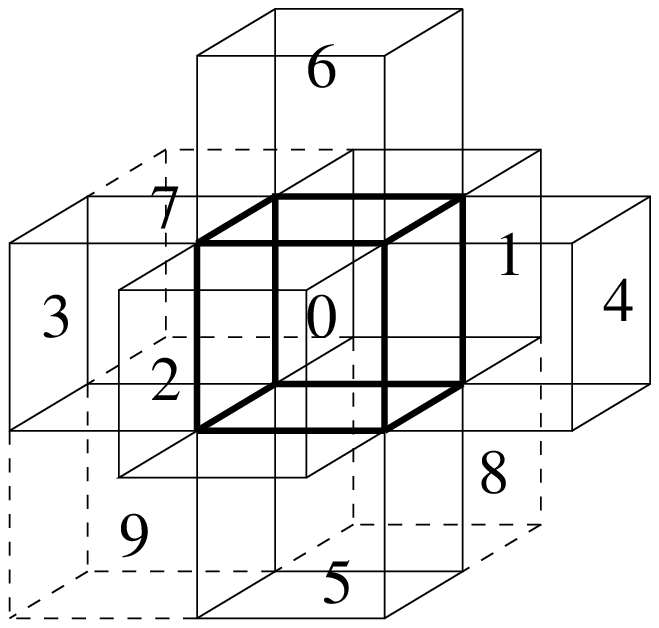,width=4.0cm,angle=270}
\end{minipage}
\end{center}
\caption[]{\footnotesize Examples of the different stencils used for 
computing derivatives are shown for the 4 different cases: a) 3 large
neighbors, b) 2 large neighbors, c) 1 large neighbor and d) 0 large 
neighbors}
\label{fig:A}
\end{figure*}

On the FTT mesh, we use the four different types of stencils shown in
Figure \ref{fig:A} when a cell has zero, one, two, or three neighbors 
that are two times larger.
These stencils involve nine neighbors of a cell and the cell itself.
Let us introduce the vector of partial derivatives of $u$ at the center of
cell $0$ 
\begin{equation}
\begin{split}
\label{eq:a1}
Du_k &= ( u, { \partial u\over\partial x},  
            { \partial u\over\partial y},  
            { \partial u\over\partial z},
            { \partial^2 u\over\partial x^2},
            { \partial^2 u\over\partial x\partial y},
            { \partial^2 u\over\partial x\partial z}, \\
&\quad
            { \partial^2 u\over\partial y^2},
            { \partial^2 u\over\partial y\partial z},
            { \partial^2 u\over\partial z^2})
\end{split}
\end{equation}
which includes the value of the function itself as the zeroth component. We
can express the values of $u_i$ in $i=1,...,9$ neighboring cells with the
second-order accurate Taylor expansion using $Du_k$. This leads to a linear
system of ten equations for the ten $Du_k$ unknowns. From this system,
$Du_k$ can be expressed as a weighted sum of the values of $u_i$ in the 
cell $0$ and its neighbors
\begin{equation}
\label{eq:a2}
Du_k = \sum_{i=0}^9 w_k^i u_i \quad .
\end{equation}
Due to the limited number of stencils encountered in the FTT structure, this
can be done once and for all and the weights can be stored in an array.

For three bigger neighbors in the positive $x$-, $y$- and
$z$-directions the weights are:
\begin{equation}
\label{eq:a3}
\begin{split}
w_1 &= {1\over 52\Delta}(-48,-14,54,-1,15,\\
&\qquad \qquad -1,15,-11,-11,2),\\
w_2 &= {1\over 52\Delta}(-48,-1,15,-14,54,\\
&\qquad \qquad -1,15,-11,2,-11),\\
w_3 &= {1\over 52\Delta}(-48,-1,15,-1,15,\\
&\qquad \qquad -14,54,2,-11,-11),\\
w_4 &= {1\over 26\Delta^{2}}(-4,14,-2,1,-15,\\
&\qquad \qquad 1,-15,11,11,-2), \\
w_5 &= {1\over \Delta^{2}}(1,0,-1,0,-1,0,0,1,0,0), \\
w_6 &= {1\over \Delta^{2}}(1,0,-1,0,0,0,-1,0,1,0), \\
w_7 &= {1\over 26\Delta^{2}}(-4,1,-15,14,-2, \\
&\qquad \qquad 1,-15,11,-2,11), \\
w_8 &= {1\over \Delta^{2}}(1,0,0,0,-1,0,-1,0,0,1), \\
w_9 &= {1\over 26\Delta^{2}}(-4,1,-15,1,-15, \\
&\qquad \qquad 14,-2,-2,11,11).
\end{split}
\end{equation}
For two bigger neighbors in the positive $x$- and $y$-directions the 
weights are:
\begin{equation}
\label{eq:a4}
\begin{split}
w_1 &= {1\over 56\Delta}(-48,-15,57,-1,15, \\
&\qquad \qquad -2,14,-12,-11,3),\\
w_2 &= {1\over 56\Delta}(-48,-1,15,-15,57, \\
&\qquad \qquad -2,14,-12,3,-11),\\
w_3 &= {1\over 2\Delta}(0,0,0,0,0,-1,1,0,0,0), \\
w_4 &= {1\over 28\Delta^{2}}(-8,15,-1,1,-15,\\
&\qquad \qquad 2,-14,12,11,-3), \\
w_5 &= {1\over \Delta^{2}}(1,0,-1,0,-1,0,0,1,0,0), \\
w_6 &= {1\over \Delta^{2}}(1,0,-1,0,0,0,-1,0,1,0), \\
w_7 &= {1\over 28\Delta^{2}}(-8,1,-15,15,-1,\\
&\qquad \qquad 2,-14,12,-3,11), \\
w_8 &= {1\over \Delta^{2}}(1,0,0,0,-1,0,-1,0,0,1), \\
w_9 &= {1\over \Delta^{2}}(-2,0,0,0,0,1,1,0,0,0).
\end{split}
\end{equation}
For one big neighbor in the positive $x$ direction the weights are:
\begin{equation}
\label{eq:a5}
\begin{split}
w_1 &= {1\over 30\Delta}(-24,-8,30,-1,7,-1,7,-6,-6,2),\\
w_3 &= {1\over 2\Delta}(0,0,0,-1,1,0,0,0,0,0), \\
w_3 &= {1\over 2\Delta}(0,0,0,0,0,-1,1,0,0,0), \\
w_4 &= {1\over 15\Delta^{2}}(-6,8,0,1,-7,1,-7,6,6,-2), \\
w_5 &= {1\over \Delta^{2}}(1,0,-1,0,-1,0,0,1,0,0), \\
w_6 &= {1\over \Delta^{2}}(1,0,-1,0,0,0,-1,0,1,0), \\
w_7 &= {1\over \Delta^{2}}(-2,0,0,1,1,0,0,0,0,0), \\ 
w_8 &= {1\over \Delta^{2}}(1,0,0,0,-1,0,-1,0,0,1), \\ 
w_9 &= {1\over \Delta^{2}}(-2,0,0,0,0,1,1,0,0,0).
\end{split}
\end{equation}
For zero big neighbors the mixed second order derivatives are given 
by the same weights as for the other three cases, while all other
derivatives are given by the standard formulas for central
differences on a uniform grid.
Equation (\ref{eq:13}) then can be expressed as
\begin{equation}
Du_4 + Du_7 + Du_9 + F(u_0) = 0 \quad .
\end{equation}

\section{Appendix: Conformal factor of two time-symmetric black holes.}
\label{ap:B}
A solution for the conformal factor of two
time-symmetric black holes \cite{bib:19}, \cite{bib:9} can be written in the
Cartesian coordinates as \cite{bib:15}
\begin{equation}
\label{eq:b1}
\phi({\bf r}) = 1 + \sum_{n=1}^\infty \left( F_1^{n} + F_2^{n} \right)
\quad ,
\end{equation}
with 
\begin{equation}
\begin{split}
F_1^n &= \begin{cases} F_1^{n-1} \left( R_1 / \rho_{11}^{n-1} \right) 
&\text{for $n$ odd};\\
F_1^{n-1} \left( R_2 / \rho_{12}^{n-1} \right)~, &\text{for $n$ even}; \\
1~, &\text{for $n=0$}, \end{cases} \\
F_2^n &= \begin{cases}  F_2^{n-1} \left( R_2 / \rho_{21}^{n-1}
\right)~, & \text{for $n$ odd}; \\
F_2^{n-1} \left( R_1 / \rho_{22}^{n-1} \right)~, &\text{for $n$ even}; \\
1~, &\text{for $n=0$}, \end{cases}
\end{split}
\end{equation}
where
\begin{equation}
\label{eq:b3}
\rho_{\alpha\delta}^{n-1} = 
\vert {\bf x}^{n-1}_\alpha - {\bf r}_\delta \vert~, 
\end{equation}
$\alpha=1,2$, $\delta=1,2$, ${\bf r}_\delta$ are the positions of the
centers of black hole throats, $R_\delta$ are the throat radii,
\begin{equation}
\label{eq:b4}
\begin{split}
{\bf x}_1^n &= \begin{cases} {\bf J}_1({\bf x}_1^{n-1})~, 
&\text{for $n$ odd}; \\
                        {\bf J}_2({\bf x}_1^{n-1})~, 
&\text{for $n$ even}; \\
                        {\bf r}~,                    &\text{for $n=0$}, 
\end{cases} \\
{\bf x}_2^n &= \begin{cases} {\bf J}_2({\bf x}_2^{n-1})~, 
&\text{for $n$ odd}; \\
                        {\bf J}_1({\bf x}_2^{n-1})~, 
&\text{for $n$ even}; \\
                        {\bf r},                    &\text{for $n=0$}, 
\end{cases} 
\end{split}
\end{equation}
and 
\begin{equation}
\label{eq:b5}
{\bf J}_\delta ({\bf x})  
     =  \left( R_\delta^2 \over  \vert {\bf x} - {\bf r}_\delta \vert^2
\right) 
\left( {\bf x} - {\bf r}_\delta \right) + {\bf r}_\delta \quad .
\end{equation}
\end{multicols}
\newpage
\begin{multicols}{2}

\end{multicols}

\end{document}